\documentclass{13086} 
\usepackage{graphicx}
\usepackage{savesym}
\savesymbol{bibfont}
\usepackage{natbib}
\restoresymbol{NB}{bibfont}
\usepackage{amsmath}
\savesymbol{iint}
\usepackage{txfonts}
\restoresymbol{TXF}{iint}
\usepackage{amssymb}
\bibpunct{(}{)}{;}{a}{}{,}
\begin{document}
\title{The magnetic precursor of L1448-mm: Excitation differences between ion and 
neutral fluids}

\author{J. F. Roberts
	\inst{1,2}
	\and
	I. Jim{\'e}nez-Serra
	\inst{3,4}
	\and
	 J. Mart{\'i}n-Pintado
	 \inst{1}
	 \and 
	 S. Viti
	 \inst{2}
	 \and
	 A. Rodr{\'i}guez-Franco
	 \inst{1,5}
	 \and
	  A. Faure
	  \inst{6} 
	  \and
	  J. Tennyson
	  \inst{2}	}

   \offprints{robertsj@inta.es}

\institute{
Centro de Astrobiolog\'{i}a (CSIC-INTA),
Ctra de Torrej\'{o}n a Ajalvir, km 4,
28850 Torrej\'{o}n de Ardoz, Madrid, Spain\\
\email{robertsj@inta.es}
\and
Department of Physics \& Astronomy, University College London, Gower Street,
London WC1E 6BT, UK\\
\and
School of Physics and Astronomy, University of Leeds, Leeds, LS2 9JT, UK\\
\and
Harvard-Smithsonian Astrophysical Observatory, 60 Garden Street, Cambridge, MA 02138, USA \\
\and
Escuela Universitaria de \'{O}ptica, Departamento de Matem\'{a}tica Aplicada (Biomatem\'{a}tica), Universidad Complutense de Madrid, 
Avda. Arcos de Jal\'{o}n s/n, E-28037 Madrid, Spain\\ 
\and
Laboratoire d'Astrophysique de Grenoble, UMR 5571-CNRS, Universit\'e Joseph
Fourier, Grenoble, France\\
}

   \date{Received August 7, 2009; accepted February 1, 2010}

% \abstract{}{}{}{}{} 
% 5 {} token are mandatory
 
  \abstract
{Shock modelling predicts an electron density enhancement 
within the magnetic precursor of  C-shocks. Previous observations of SiO,
H$^{13}$CO$^{+}$, HN$^{13}$C and H$^{13}$CN toward the young 
L1448-mm outflow showed an over-excitation of the ion fluid that was 
attributed to an electron density enhancement in the precursor.}
{We re-visit this interpretation and test if it still holds when we consider 
different source morphologies  and kinetic temperatures for the observed molecules.
 To do this, we use updated collisional coefficients of HN$^{13}$C and SiO with electrons in our excitation model.
We also aim to give some insight on the spatial extent of the 
electron density enhancement around L1448-mm.}
{We estimate the opacities of H$^{13}$CO$^+$ and HN$^{13}$C by observing 
the $J$=3$\rightarrow$2 lines of rarer isotopologues.
 To model the excitation of the molecules, we use the large velocity gradient (LVG) approximation with updated collisional coefficients to i) re-analyse the observations toward the positions where the over-excitation of H$^{13}$CO$^+$ has previously been observed [i.e. toward L1448-mm at offsets (0,0) and (0,-10)], and ii) to investigate if the electron density enhancement is still required for the cases of extended and compact emission, and for kinetic temperatures of up to 400~K.
We also report several lines of SiO, HN$^{13}$C and H$^{13}$CO$^+$ 
toward new positions around this outflow, to investigate the spatial extent of 
the over-excitation of the ions in L1448-mm.}
{From the isotopologue observations, we find that the emission of 
H$^{13}$CO$^+$ and HN$^{13}$C from the precursor is optically thin if this 
emission is extended.  
Using the new collisional coefficients, an electron density enhancement is still needed to explain the excitation of H$^{13}$CO$^+$ for extended emission and for gas temperatures of $\leq 400$~K toward L1448-mm (0,-10), and possibly also toward L1448-mm (0,0).  For compact emission the data cannot be fitted.
We do not find any evidence for the 
over-excitation of the ion fluid toward the newly observed positions around
L1448-mm.}
{The observed line emission of SiO, H$^{13}$CO$^{+}$ and HN$^{13}$C  toward L1448-mm (0,0) and (0,-10) 
is consistent with an electron density enhancement in the precursor
component, if this emission is spatially extended. 
This is also true for the case of 
high gas temperatures ($\leq$400$\,$K) toward the (0,-10) offset.
The electron density enhancement seems to be restricted to the southern, 
redshifted lobe of the L1448-mm outflow. Interferometric images of the 
line emission of these molecules are needed to confirm the spatial extent
of the over-excitation of the ions and thus, of the electron density 
enhancement in the magnetic precursor of L1448-mm.}


   \authorrunning{Roberts et al.}
   \titlerunning{The magnetic precursor of L1448-mm}
   \maketitle
%
%________________________________________________________________

\section{Introduction}

C-shock waves associated with molecular outflows  are belived to develop a 
 thin region of enhanced fractional ionisation 
known as the `magnetic precursor' \citep{Draine80,Flower96,Flower03}.
Inside this region, the magnetic field is gradually
compressed, forcing the ions to stream through the neutral gas, accelerating, 
compressing and heating this fluid before the neutral one.
The subsequent ion-neutral velocity decoupling leads to the sputtering
of dust grains, injecting large amounts of molecular material into the gas 
phase 
%(see e.g. Caselli et al. 1997; Schilke et al. 1997; Jimenez-Serra et al. 2008).
 \citep[e.g.][]{Caselli97,Schilke97,J-S08}
 The electron density is predicted to be enhanced within the magnetic precursor by
 a factor of $\sim100$ 
due to the  fluorescence UV photons generated after the
collisional excitation of H$_2$ molecules \citep{Flower96,Flower03}. 

It has been proposed that a narrow (line-width $\sim 0.5$~km~s$^{-1}$) 
and very low velocity component of SiO detected towards the very young 
\object{L1448-mm} outflow, is the signature of the interaction of the magnetic 
precursor (Jim{\'e}nez-Serra et al. \citeyear{J-S04},  hereafter JS04).
The derived fractional 
abundance of SiO is of the order $10^{-11}$
in the velocity component of the precursor,  an 
enhancement by a factor of 10 with respect to 
the upper limits measured in the 
quiescent gas of dark clouds  
\citep[$\leq$10$^{-12}$ in L183 and L1448;][]{Ziurys89,RT07}.
%($\leq$10$^{-12}$ in L183 and L1448; Ziurys et al. 1989; Requena-Torres et al. 2007).
Although the origin of this narrow SiO emission is still unclear, 
it has been suggested that the sputtering of dust grains at the precursor 
stage, is efficient enough to inject a considerable fraction
of the grain mantles into the gas phase. As a consequence, bright and 
narrow SiO line emission is expected to arise from material whose 
central radial velocities are very similar
to that of the ambient cloud \citep{J-S08,J-S09}.

In correlation with the detection of narrow SiO emission, 
Jim{\'e}nez-Serra et al. (\citeyear{J-S06}, hereafter JS06) reported 
differences in the excitation of molecular ions, such as
H$^{13}$CO$^{+}$, with respect to neutral molecules, such as HN$^{13}$C and H$^{13}$CN,
in the precursor component. In particular, the high-$J$ line emission of 
H$^{13}$CO$^+$ is substantially brighter than that of HN$^{13}$C toward 
those regions where the precursor has been detected. JS06 proposed that 
these differences, which cannot be accounted for 
by considering {\it only} molecular excitation by collisions with H$_2$ and 
a single H$_2$ density of few $\times$10$^5$$\,$cm$^{-3}$ for the molecular gas, 
could be produced by the {\it selective} excitation of molecular ions 
by collisions with electrons 
 within the precursor.
This study 
established that the over-excitation of the ions is consistent with an 
electron density enhancement by a factor of $\sim$500 in the precursor component  toward L1448-mm (0,0) and (0,-10),
which is similar to that predicted by C-shock modelling at this 
 dynamical time
\citep{Flower03}. 

However, the observed differences in excitation between the ion and the neutral molecules
could be alternatively explained by opacity effects. Indeed, \citet{Frerking79}
and \citet{Cernicharo84} showed anomalies in the {\it large-scale} 
line intensity emission 
of the hyperfine components of HNC and HCN toward the molecular dark clouds
TMC-1 and L134N. The equal intensities measured for these components could be
produced by the absorption of the emission arising from the dense cores by 
the less dense foreground material. This also applies to other high gas density 
tracers with high dipolar moment such as HCO$^+$, since they also show 
similar behaviors for the line intensity ratios between its isotopologue 
species \citep{Langer78}. 

JS06 derived the electron density enhancement in the precursor component
of L1448-mm by assuming that the H$^{13}$CO$^{+}$, the HN$^{13}$C
and the H$^{13}$CN emission had thin to moderate optical depths.
Unfortunately, the lack of even
rarer isotopologue observations toward this outflow 
 prevented, first, to determine the actual optical depths of this 
emission, and second, to clearly establish the origin of the over-excitation
of the ions in L1448-mm.   

In this paper we present new observations of $J$=3$\to$2 lines of the rare isotopologues 
HC$^{18}$O$^+$, H$^{13}$C$^{18}$O$^+$, 
H$^{15}$NC and H$^{15}$N$^{13}$C, toward the regions where the over-excitation 
of the ions has been reported. 
 In addition, we present new observations of the $J$=3$\to$2 lines of 
HN$^{13}$C and H$^{13}$CO$^+$ toward extra positions around the L1448-mm 
source,
 which we combine with previous observations toward these positions, 
observed by JS04, to derive the spatial extent of the 
over-excitation of H$^{13}$CO$^+$ associated with the precursor.
These observations will allow to prove that the anomalous excitation of the ions
is not due to a large-scale scattering effect, but to a real enhancement of 
the local density of electrons at the early stages of the interaction of 
very young C-shocks.

 We also re-visit the analysis of JS06, using the same excitation model as JS06 but including new collisional data 
of HNC and SiO with electrons  \citep[][respectively]{Faure07,Varambhia09}, to test if the
conclusions of JS06 still hold.  
 The excitation model employed uses the large velocity gradient (LVG) approximation.
We note that the L1448-mm outflow is likely 
 the best object where the effects of collisional 
excitation by electrons on the molecular excitation can be directly tested.  
Furthermore, in our re-analysis we fully explore the effects
of source morphology  and kinetic gas temperature on our results.

In Sections~\ref{obs_description} and \ref{results}, we present the 
observations carried out with the JCMT telescope and 
describe the results. In Section~\ref{opacities_calc}, we estimate the 
opacities of H$^{13}$CO$^+$ and HN$^{13}$C toward the positions where the 
over-excitation of the ions has been reported, assuming 
both compact and extended source emission. 
 In Section~\ref{section:excitation}, we present the re-analysis of the
LVG calculations of JS06 using the new collisional coefficients,  considering different source morphologies and temperatures up to 400~K.
 In this section we also
 analyse the data measured toward the new  
positions in L1448-mm  to provide some insight into the  spatial extent
of the 
over-excitation of H$^{13}$CO$^+$. Finally, discussion and conclusions are presented 
in  Section~\ref{obs_conclusions}.

\section{Observations}
\label{obs_description}

We observed the $J$=3$\to$2 transition of 
 the rare isotopologues HC$^{18}$O$^+$, 
H$^{13}$C$^{18}$O$^+$ 
and H$^{15}$N$^{13}$C
toward L1448-mm at offsets (0,0) 
and (0,-10),
 and H$^{15}$NC toward L1448-mm (0,0).
The coordinates of the central position of the L1448-mm source are 
($\alpha = 03^{h}25^{m}38^{s}.8$ , 
$\delta = 30^\circ44'05''.4$ [J2000]). 
 Finally, we measured the $J$=3$\to$2 lines of 
H$^{13}$CO$^+$ and HN$^{13}$C toward the offsets (0,20), (20,0) 
and (-20,0).

 These lines, with frequencies ranging from 248 to 267$\,$GHz, 
were observed with the 
James Clerk Maxwell Telescope (JCMT) at Mauna Kea (Hawaii), using the 
position switched observing mode and a reference (off) position of 
(800$''$,0) with respect to the central position.
The ACSIS spectrometer provided a spectral resolution of 31~kHz, which 
corresponds to a velocity resolution of $\sim 0.04$~km~s$^{-1}$ for the 
A3 receiver. The system temperatures ranged from 400~K to 700~K.

The $J$=3$\to$2 lines of H$^{13}$CO$^+$ and HN$^{13}$C (at $\sim$260$\,$GHz),
observed with the IRAM 30m telescope at Pico Veleta (Spain) 
toward L1448-mm (0,0) and (0,-10), were taken from JS06. 
The spectra of the $J$=1$\to$0  
transitions of H$^{13}$CO$^+$ and HN$^{13}$C, and of the $J$=2$\to$1 
transition of SiO (at $\sim 87$~GHz) toward L1448-mm 
(0,20), (20,0) and (-20,0), were also observed with this telescope,
 and have been previously published in JS04.
 See JS06 and JS04 for a full description of these observations.

 Table~\ref{beamsize} summarises the beam efficiencies and sizes, for all of the
observations analysed in this paper.
All the intensities were calibrated in units of antenna temperature ($T_A^*$).

%TABLE 1-----------------------------------------------------------------------------------
\begin{table*}
\caption{ Summary of beam efficiencies and FWHM beam sizes for all the observations
analysed in this paper.}
\begin{tabular}{lcccccc} \hline
Lines & Offset ($''$) & Frequency (GHz) & Telescope & $\eta$ & beam size ($''$)&Reference \\ \hline
HC$^{18}$O$^+$(3-2), H$^{13}$C$^{18}$O$^+$(3-2),& (0,0), (0,-10) & $248-267$ & JCMT & 0.69 & 21 & This work \\
H$^{15}$N$^{13}$C(3-2), H$^{15}$NC(3-2) & & & & & &  \\ \hline
H$^{13}$CO$^+$(3-2), HN$^{13}$C(3-2)&(0,20), (20,0), (-20,0) &  $\sim 260$ & JCMT & 0.69 & 21 & This work \\    \hline
SiO(2-1), HN$^{13}$C(1-0), &(0,0), (0,-10) & $\sim 90$ & IRAM & 0.82 & 28 & JS06 \\ 
H$^{13}$CO$^+$(1-0) & & & & & & \\ \hline
SiO(3-2) & (0,0), (0,-10) &$\sim 130$ & IRAM & 0.74 & 19 & JS06  \\ \hline
SiO(5-4), HN$^{13}$C(3-2), &(0,0), (0,-10) &$217-260$&IRAM &0.52 & 11  & JS06 \\
H$^{13}$CO$^+$(3-2) & & & & & & \\ \hline
HN$^{13}$C(4-3), H$^{13}$CO$^+$(4-3) & (0,0), (0,-10) &$\sim 350$& JCMT & 0.63 & 14  & JS06 \\ \hline
H$^{13}$CO$^+$(1-0), HN$^{13}$C(1-0)& (0,20), (20,0), (-20,0) & $\sim87$ & IRAM & 0.82 & 28 & JS04 \\
\hline

\end{tabular}
\label{beamsize}

\end{table*}
%--------------------------------------------------------------------------------------------

\section{Results}
\label{results}

%______________________________________________ Figure 1
   \begin{figure}
   \includegraphics[angle=0, width=0.5\textwidth]{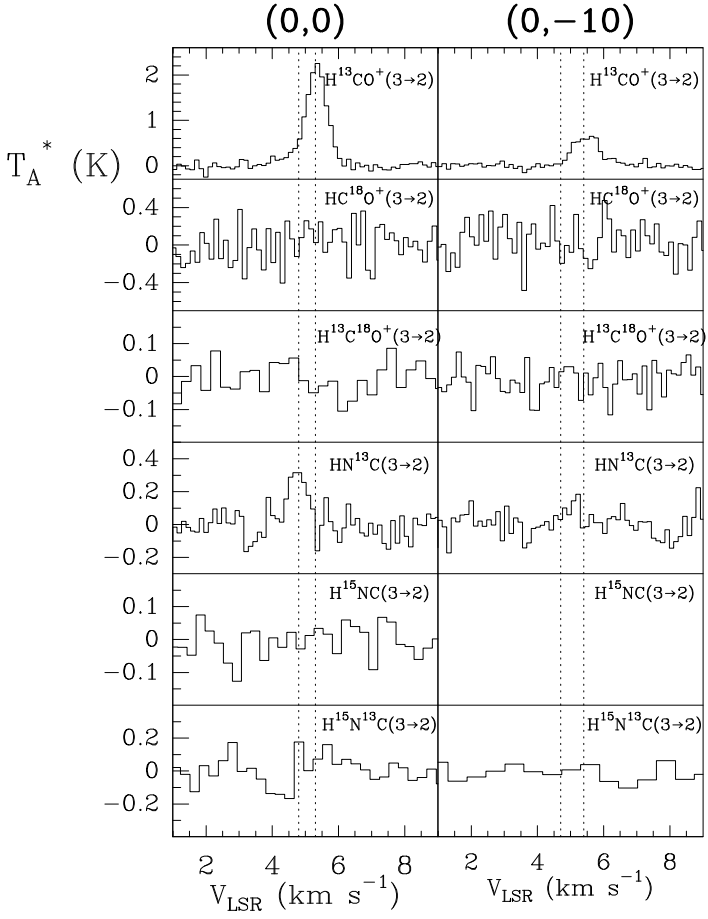}
\caption{Observations of the  $J$=3$\rightarrow$2 line of several 
isotopologue species of HCO$^+$ and HNC toward the L1448-mm (0,0)
and (0,-10) positions. Offsets, in arcseconds, are shown in the upper 
part of the columns. The vertical dotted lines show the ambient gas 
centred at 4.7~km~s$^{-1}$, and the precursor component at 
5.2 and 5.4~km~s$^{-1}$ toward L1448-mm (0,0) and (0,-10), respectively (see JS06).} 
\label{opacity}
    \end{figure}
%__________________________________________________

%TABLE 2-------------------------------------------------------------------

\begin{table*}
\begin{small}
\caption{Observed parameters of the $J$=3$\to$2 lines of 
H$^{13}$CO$^+$, HC$^{18}$O$^+$, H$^{13}$C$^{18}$O$^+$, HN$^{13}$C, 
H$^{15}$NC and H$^{15}$N$^{13}$C toward L1448-mm (0,0) and (0,-10). }  

\begin{center}
\begin{tabular}{lcccccc} \hline 
&  \multicolumn{3}{c}{(0,0)} & \multicolumn{3}{c}{(0,-10)}  \\ \cline{2-4} \cline{5-7} 

& $V_{LSR}$& $\Delta v$ & $T_A^*$ & $V_{LSR}$& $\Delta v$ & $T_A^*$  \\
 Line &  (km s$^{-1}$) & (km s$^{-1}$) & (K) &  (km s$^{-1}$) & (km s$^{-1}$) & (K) \\ \hline
H$^{13}$CO$^+$$^{\mathrm{a}}$ & 4.5(2) & 0.9(3) & 0.23(9) &  $\sim$4.7 & ... & $\leq$0.168  \\
& 5.35(1) & 0.71(3) & 2.21(9) &  5.52(4) & 0.90(8) & 0.65(6) \\
HC$^{18}$O$^+$ & $\sim 4.7$ & ... & $\leq 0.195$$^{\mathrm{b}}$ & $\sim4.7$ & ... & $\leq0.195$  \\
&  $\sim 5.2$ & ... & $\leq 0.195$ &  $\sim5.2$ & ... & $\leq0.195$\\
H$^{13}$C$^{18}$O$^+$ & $\sim 4.7$ & ... & $\leq 0.090$&$\sim 4.7$ & ... & $\leq0.087$\\
& $\sim 5.2$ & ... & $\leq 0.090$&$\sim 5.2$ & ... & $\leq0.087$\\
HN$^{13}$C$^{\mathrm{a}}$ &4.75(5)&0.6(1)&0.35(9)&$\sim 4.7$&...& $\leq0.261$\\
&$\sim5.2$&...&$\leq0.237$&$\sim 5.2$&...& $\leq0.261$\\
H$^{15}$NC & $\sim4.7$&...& $\leq0.093$& ...& ... & ... \\
& $\sim5.2$&...& $\leq0.093$& ...& ... & ... \\
H$^{15}$N$^{13}$C&$\sim4.7$ & ... & $\leq0.198$ &  $\sim4.7$ & ... & $\leq 0.207$ \\
&$\sim5.2$ & ... & $\leq0.198$ &  $\sim5.2$ & ... & $\leq 0.207$\\ \hline
\end{tabular}
\end{center}
\label{Opacity-par}

\begin{list}{}{}
\item[$^{\mathrm{a}}$] These molecular line transitions were taken from JS06.
\item[$^{\mathrm{b}}$] The upper limits of the line intensities correspond 
to the $3\sigma$ noise level.
\end{list}

\end{small}
\end{table*}%

%----------------------------------------------------------------------------------

The line profiles of the $J$=3$\to$2 transitions of the rare isotopologues 
of HCO$^+$ and HNC observed toward L1448-mm (0,0) and (0,-10), 
are shown in Figure~\ref{opacity}. The $J$=3$\to$2 lines of H$^{13}$CO$^+$ and 
HN$^{13}$C  from JS06 are also plotted. The observed parameters of this 
emission 
are given in Table~\ref{Opacity-par}. Where detected, the line intensities 
are well above the 
$3\sigma$ noise level.
 These data are used to calculate the upper limits on the opacity of the 
H$^{13}$CO$^{+}$ and HN$^{13}$C emission in Section~\ref{opacities_calc}.

The H$^{13}$CO$^{+}$ molecular line emission
toward L1448-mm (0,0) was fitted with a double gaussian line profile,
whose radial velocity components are centred at 
$V_{LSR}$=4.7$\,$km$\,$s$^{-1}$ (the ambient gas), and at 
$V_{LSR}$=5.2$\,$km$\,$s$^{-1}$ (the magnetic precursor component). 
These two velocity components were already reported by 
JS04 and JS06.

Toward L1448-mm (0,-10), the H$^{13}$CO$^{+}$  $J$=3$\rightarrow$2 emission
shows a single  peak
centred at the precursor component
with $V_{LSR}$=5.4$\,$km$\,$s$^{-1}$. The 
HN$^{13}$C $J$=3$\rightarrow$2 line, which is only detected toward
L1448-mm (0,0),  has a single peak at the ambient gas velocity, $V_{LSR}$=4.7$\,$km$\,$s$^{-1}$.
The  
$J$=3$\to$2 lines of the rare isotopologues HC$^{18}$O$^+$, 
H$^{13}$C$^{18}$O$^+$, H$^{15}$NC and H$^{15}$N$^{13}$C were not detected 
 above the $3\sigma$ limit  (see Table$\,$2).

%______________________________________________ Figure 2
   \begin{figure}
   \centering
   \includegraphics[angle=0, width=0.5\textwidth]{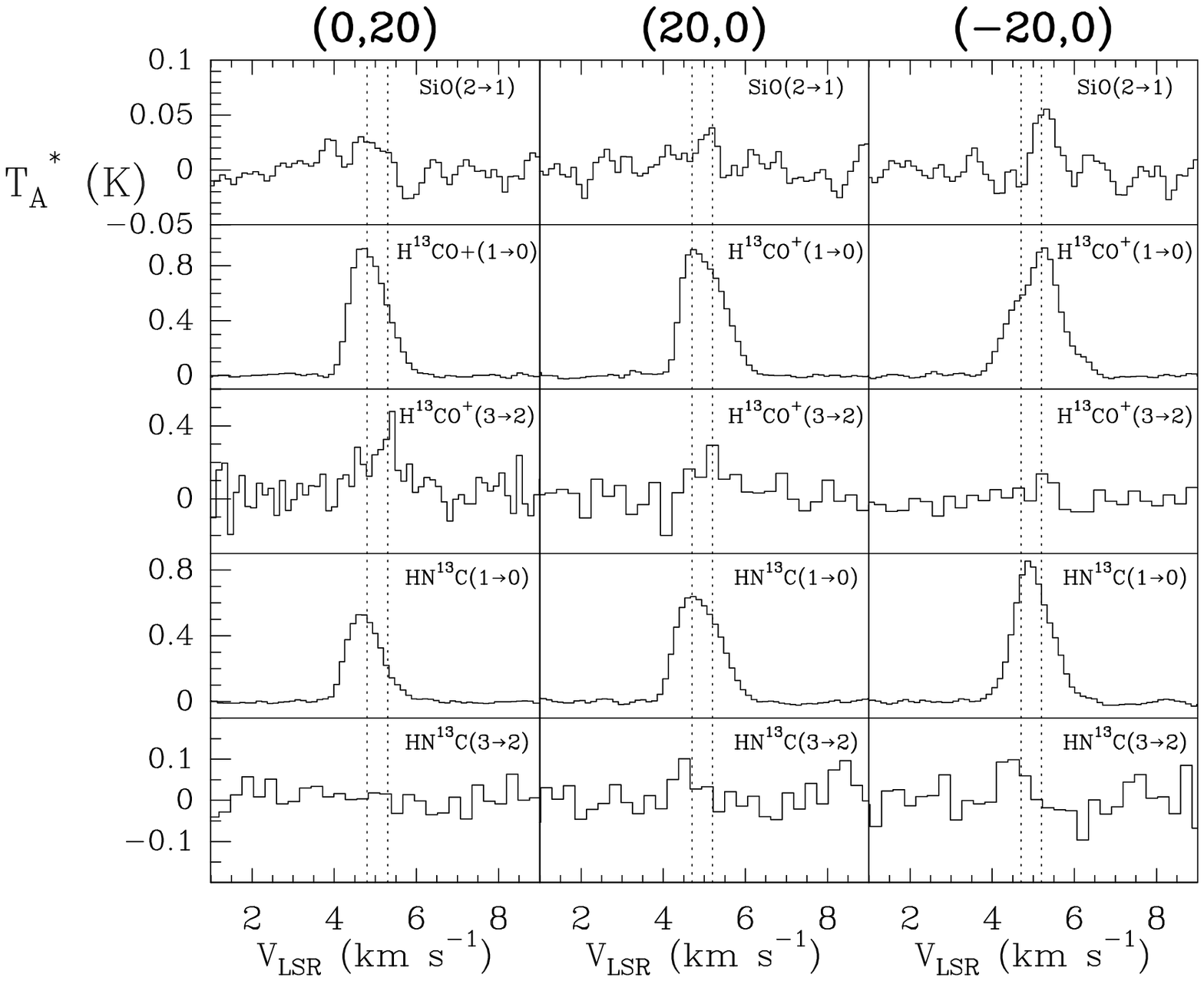}
\caption{Observations of the SiO $J$=2$\to$1 line emission and of 
the $J$=3$\to$2 and 1$\to$0 transitions of H$^{13}$CO$^+$ and HN$^{13}$C
toward the L1448-mm (0,20), (20,0) and (-20,0) positions. Offsets, in arcseconds,
are shown in the upper part of the columns. The vertical dotted lines show 
the ambient gas at 4.7~km~s$^{-1}$ and the precursor component at 
5.2~km~s$^{-1}$. }
\label{excitation}
    \end{figure}
%__________________________________________________

%TABLE 3------------------------------------------------------------------------------------

\begin{table*}
\begin{small}
\caption{Observed parameters of the $J$=2$\to$1 line of SiO, and of the
$J$=3$\to$2  and $J$=1$\to$0 transitions of H$^{13}$CO$^+$ and HN$^{13}$C 
toward L1448-mm (0,20), (20,0) and (-20,0).} 
\begin{center}
\begin{tabular}{lccccccccc} \hline 
&  \multicolumn{3}{c}{(0,20)} & \multicolumn{3}{c}{(20,0)} & \multicolumn{3}{c}{(-20,0)} \\ \cline{2-4} \cline{5-7} \cline{8-10}
& $V_{LSR}$& $\Delta v$ & $T_A^*$ &  $V_{LSR}$& $\Delta v$ & $T_A^*$ &  $V_{LSR}$& $\Delta v$ & $T_A^*$ \\
 Line &  (km s$^{-1}$) & (km s$^{-1}$) & (K) &  (km s$^{-1}$) & (km s$^{-1}$) & (K) &  (km s$^{-1}$) & (km s$^{-1}$ & (K) \\ \hline

SiO($2\to1$)$^{\mathrm{a}}$ & $\sim 4.7$& \ldots & $\leq0.027$$^{\mathrm{b}}$ & $\sim 4.7$ & \ldots &$\leq 0.030 $&  $\sim 4.7$ & \ldots &$\leq 0.036$ \\
& $\sim 5.2 $ & \ldots & $\leq 0.027$& 5.07(6) & 0.5(1) & 0.04(1) & 5.29(4) & 0.51(7) & 0.06(1) \\

H$^{13}$CO$^+$($1\to 0$)$^{\mathrm{a}}$ & 4.597(3) & 0.625(7) & 0.75(2) & 4.645(4) & 0.590(9) & 0.74(2) & 4.606(3) & 0.87(2) & 0.49(3) \\
&5.126(0) & 0.82(1) & 0.57(2) & 5.254(0) & 0.89(1) & 0.68(2) & 5.318(1) & 0.72(1) & 0.85(3) \\

H$^{13}$CO$^+$($3\to 2$) & 4.5(1) & 0.5(3) & 0.20(6) & $\sim 4.7$ & \ldots & $\leq 0.14$ & $\sim 4.7$ & \ldots & $\leq 0.09$ \\
&  5.33(7) & 0.7(2) & 0.36(5) & 5.2(1) & 1.0(3) & 0.22(5) & 5.3(2) & 0.4(2) & 0.14(3) \\

HN$^{13}$C($1\to 0$)$^{\mathrm{a}}$ & 4.625(6) & 0.80(1) & 0.52(1) & 4.6(1) & 0.8(1) & 0.57(1) & 4.89(1) & 0.86(2) & 0.84(1) \\
& 5.224(2) & 0.80(5) & 0.15(1) & 5.2(1) & 0.9(1) & 0.40(1) & 5.58(4) & 0.63(6) & 0.21(1) \\

HN$^{13}$C($3\to 2$) & $\sim 4.7$ & \ldots & $\leq 0.07$ & $\sim 4.7$ & \ldots & $\leq 0.075$ & $4.5$(2) & 0.7(2) & $0.09(2)$ \\
&$\sim 5.2$ & \ldots & $\leq 0.07$ & $\sim 5.2$ & \ldots & $\leq 0.075$ & $\sim 5.2$ & \ldots & $\leq 0.07$  \\ \hline

\end{tabular}
\end{center}
\label{Obs-par}

\begin{list}{}{}
\item[$^{\mathrm{a}}$] These molecular line transitions were taken from JS04.
\item[$^{\mathrm{b}}$] The upper limits of the line intensities correspond 
to the $3\sigma$ noise level.
\end{list}

\end{small}
\end{table*}%

%------------------------------------------------------------------------------------------

The profiles of the SiO $J$=2$\to$1 lines, and of the $J$=3$\to$2 and 1$\to$0 
transitions of H$^{13}$CO$^+$ and HN$^{13}$C toward the L1448-mm 
(0,20), (20,0) and (-20,0) offsets, are shown in Figure~\ref{excitation}. 
The observed parameters are given in Table~\ref{Obs-par}, 
  with all detections above the $3 \sigma$ noise level.

Narrow,  single-peaked SiO $J$=2$\to$1
emission is detected toward L1448-mm (-20,0), 
and very faintly toward L1448-mm (20,0). These lines are 
centred at  the velocity of the precursor component
($\sim 5.2$~km~s$^{-1}$), and show line-widths of $\sim 0.5$~km~s$^{-1}$  (Table$\,$3). 
 The line profiles of H$^{13}$CO$^+$ $J$=1$\rightarrow$0 and HN$^{13}$C 
$J$=1$\rightarrow$0 are fitted by double 
Gaussian profiles, with line widths of $0.6-0.8$~km~s$^{-1}$ for each component.
Toward L1448-mm (0,20) and (20,0), the H$^{13}$CO$^+$ $J$=1$\rightarrow$0 lines show 
their maximum emission at the ambient velocity component with 
V$_{LSR}$=4.7$\,$km$\,$s$^{-1}$. 
However, toward L1448 (-20,0) where the detection of narrow SiO is more 
evident, 
H$^{13}$CO$^+$ $J$=1$\rightarrow$0 peaks at the precursor component with 
V$_{LSR}$=5.2$\,$km$\,$s$^{-1}$. 
The  peak emission of the HN$^{13}$C $J$=1$\rightarrow$0 lines are all centred 
at the velocity of the ambient gas ($V_{LSR}$$\sim$4.7$\,$km$\,$s$^{-1}$), 
except toward L1448-mm (-20,0) where this emission is red-shifted to
$\sim$4.9$\,$km$\,$s$^{-1}$  (Table$\,$3).  

The $J$=3$\to$2 lines of H$^{13}$CO$^+$ show single peaked line profiles 
centred at $\sim 5.2$~km~s$^{-1}$ toward L1448-mm (20,0) and (-20,0), while 
toward L1448-mm (0,20), a double gaussian line profile can roughly be fitted.
 This behaviour was also reported by JS06 toward L1448-mm (0,0), where the
$J$=3$\to$2 and $J$=4$\to$3 lines of H$^{13}$CO$^+$ also show faint emission arising from
the ambient gas. As shown by JS06, this emission is consistent with the excitation 
of H$^{13}$CO$^+$ by collisions with only H$_2$, assuming an H$_2$ density of 
$\sim$10$^5$~cm$^{-3}$.

 Toward (20,0) and (-20,0), the detection of H$^{13}$CO$^+$ emission only
in the precursor component suggests that the ions could have already been 
accelerated within the shock to larger red-shifted velocities. This is more 
apparent toward (-20,0) where all molecular lines have their peak emission 
slightly red-shifted to larger velocities compared to other positions.    
As proposed by JS06, this velocity shift could be understood as an 
evolutionary effect in which the molecular material toward L1448-mm (-20,0) 
is at a later  dynamical time within the C-shock. 
This is also supported by the clear detection of narrow SiO toward this 
position. We finally note that no signal above the $3\sigma$ noise level was 
detected for the HN$^{13}$C $J$=3$\rightarrow$2 lines at any position in 
L1448-mm  for the precursor component.
In Section~\ref{new-pos-obs}, all these data will be analysed to establish 
the spatial extent of the over-excitation of H$^{13}$CO$^+$ in L1448-mm, 
and its implications for the magnetic precursor scenario. 

\section{Calculation of opacities}
\label{opacities_calc}

In order to rule out the possibility that the over-excitation of 
H$^{13}$CO$^+$ is due to large opacity effects, we need to prove that the 
emission lines of H$^{13}$CO$^+$ and HN$^{13}$C reported by JS06 
are optically thin. In addition, low optical depths associated with the
H$^{13}$CO$^+$ and HN$^{13}$C emission toward L1448-mm will fully justify 
the use of the LVG approximation as the molecular excitation model 
to derive 
the electron density enhancement predicted at the precursor stage.

%TABLE 4---------------------------------------------------------------------------------

\begin{table}
\caption{Isotopic abundance ratios for the local ISM taken from \cite{Wilson99}.}
\begin{center}
\begin{tabular}{c|c}
 Species &  Ratio ($R$) \\ \hline 
$\frac{\text{HC}^{18}\text{O}^+}{\text{H}^{13}\text{CO}^+} =  \frac{[^{12}\text{C}]}{[^{13}\text{C}]} \times  \frac{[^{18}\text{O}]}{[^{16}\text{O}]}  $ & $\frac{69}{577}$\\
$\frac{\text{H}^{13}\text{C}^{18}\text{O}^+}{\text{H}^{13}\text{CO}^+} =  \frac{[^{18}\text{O}]}{[^{16}\text{O}]}$ & $\frac{1}{577}$ \\
$\frac{\text{H}^{15}\text{NC}}{\text{HN}^{13}\text{C}} = \frac{[^{15}\text{N}]}{[^{14}\text{N}]} \times  \frac{[^{12}\text{C}]}{[^{13}\text{C}]}  $ & $\frac{69}{388}$ \\
$\frac{\text{H}^{15}\text{N}^{13}\text{C}}{\text{HN}^{13}\text{C}} = \frac{[^{15}\text{N}]}{[^{14}\text{N}]}$ & $\frac{1}{388} $ \\

\end{tabular}
\end{center}
\label{R}
\end{table}%
%-------------------------------------------------------------------------------------------

 We can estimate the optical depths of the H$^{13}$CO$^+$ and HN$^{13}$C $J=3\to2$ lines, $\tau_{32}$,  toward L1448-mm (0,0) and (0,-10) using the observed isotopic line intensity ratios \citep{Choi93}: 
\begin{equation}
\frac{T_{\text{B,rare}}} {T_{\text{B},32}} = \frac{1-e^{-R\tau_{32}}}{1-e^{-\tau_{32}} }.
\label{opacity_eqn}
\end{equation}

This equation assumes equal excitation temperature for  both isotopologues.
Here, $T_{\text{B},32}$ is the {\it brightness} temperature of the H$^{13}$CO$^+$ $J$=3$\rightarrow$2 or HN$^{13}$C $J$=3$\rightarrow$2 line, and $T_{\text{B},rare}$ is the brightness temperature of the $J=3\to2$ line of the corresponding rarer isotopologue (either HC$^{18}$O$^+$ or H$^{13}$C$^{18}$O$^+$ for H$^{13}$CO$^+$, or H$^{15}$NC or H$^{15}$N$^{13}$C for HN$^{13}$C).
$R$ is  the typical isotopic abundance ratio
observed in the local interstellar medium, values for which are given in 
Table~\ref{R} \citep{Wilson99}.  
These values are  averages for the local ISM. 

 To estimate the brightness temperature ratios from the values of 
$T_\text{A}^*$ given in Table~\ref{Opacity-par}, we need to know the source 
morphology.
Since the morphology of the precursor component of the H$^{13}$CO$^+$ and HN$^{13}$C 
emission is
unknown, we assume two possible extreme cases: compact emission with a source  size of $5''$, and extended 
emission.

 For the source size of $5''$, the line intensities in units of 
$T_\text{A}^*$ are converted to brightness temperatures by the formula
$T_\text{B}=\frac{T_\text{A}^*}{\eta}(\theta_\text{source}^2 + \theta_\text{beam}^2)/\theta_\text{source}^2$
where $\theta_{source}$ is the source size, $\theta_\text{beam}$ is the beam size
and $\eta$ is the beam efficiency.  Values for $\theta_\text{beam}$ and $\eta$ are given in 
Table~\ref{beamsize}.
For the extended source case, the brightness temperatures are approximated by $T_\text{A}^*$.

The estimated line intensity ratios and optical depths for the 
H$^{13}$CO$^+$ and HN$^{13}$C  $J$=3$\rightarrow$2 emission are given in Tables~\ref{tab:opacities1} 
and~\ref{tab:opacities2}, respectively, for the different source morphologies 
and isotopologue ratios considered. 

 An inspection of Tables~\ref{tab:opacities1} and~\ref{tab:opacities2} clearly shows that, many line ratios are larger than unity due to  the lack of sensitivity of our observations for the rarer isotopologues. Under these conditions equation~\eqref{opacity_eqn} cannot be used to estimate the opacities.

 From Tables~\ref{tab:opacities1} and~\ref{tab:opacities2} we draw the following conclusions:

\begin{description}

\item[{\bf Ambient components:}]
For the ambient components of the H$^{13}$CO$^+$ $J$=3$\rightarrow$2 emission, the data is not sensitive enough to provide useful constraints.
For HN$^{13}$C $J$=3$\rightarrow$2 toward (0,0), assuming an extended source, the ambient component is relatively optically thin ($\tau_{32}\leq1.1$, derived from the H$^{15}$NC/HN$^{13}$C ratio). 
Since the line emission from the ambient component has indeed been observed to be extended \citep{Bachiller90,Curiel99}, 
from our data on  HN$^{13}$C $J$=3$\rightarrow$2 we conclude that the emission from this component is optically thin, as assumed by JS06.  
The ambient component will not be further considered in this paper.

\item[{\bf H$^{13}$CO$^+$ $J$=3$\rightarrow$2 precursor component toward (0,0):}]  If the source size is $\geq 5''$, this line is relatively optically thin ($\tau_{32} \leq 2.1$, using the estimate derived from the HC$^{18}$O$^+$/H$^{13}$CO$^+$ ratio).  For the extended source case, the HC$^{18}$O$^+$/H$^{13}$CO$^+$ ratio has an upper limit of 0.09, which results in a negative solution for the optical depth for the assumed value of $R$.  This is because equation~\eqref{opacity_eqn} only has a positive solution for $\tau_{32}$ for $R \leq T_{\text{B},rare}/T_{\text{B},32} \leq 1$.  
Even if we consider the lower limit of the  H$^{13}$CO$^+$ $J$=3$\rightarrow$2
line intensity,  corresponding  to the $1\sigma$ noise level, the upper limit of  $T_{\text{B},rare}/T_{\text{B},32}$ remains less than $R$. This  suggests that the actual isotopic ratios in L1448 deviate from the averaged values of $R=69/577$ in Table~\ref{R}. Significant variations in the isotopic ratios in nearby dark clouds have been  recently found. Indeed, the   $^{12}$C/$^{13}$C ratio has been estimated to vary from $\sim 40$ to $\sim 90$ within 
the local ISM \citep{Casassus05}. If we assume a extreme value 
of $^{12}$C/$^{13}$C$=40$, $R$ decreases to $\sim 0.07$ providing an 
optical depth of $\tau_{32}\lesssim0.6$.  Therefore, our 
data suggest a rather low $^{12}$C/$^{13}$C ratio and/or a high $^{16}$O/$^{18}$O 
ratio in the L1448 molecular cloud  if the emission is extended. In summary, the emission of H$^{13}$CO$^+$ $J$=3$\rightarrow$2 for the precursor toward (0,0) is optically thin even if it arises from a small source. 

\item[{\bf H$^{13}$CO$^+$ $J$=3$\rightarrow$2 precursor component toward (0,-10):}]  If the source is extended, this line is also relatively optically thin  ($\tau_{32}\leq 2.8$, derived from the HC$^{18}$O$^+$/H$^{13}$CO$^+$ ratio).  For a source size of $5''$, we can only constrain $\tau_{32}$ to $\leq 16$.

\item[{\bf HN$^{13}$C $J$=3$\rightarrow$2 precursor component toward (0,0) and (0,-10):}]  For both geometries we cannot constrain the optical depths due to the lack of sensitivity of our observations.
However, for a source of size $5''$, we find in Section~\ref{compact} that our LVG model cannot reproduce the observed intensities of HN$^{13}$C, implying that the emission should be extended.

\end{description}

%TABLE 5------------------------------------------------------------------------------

\begin{table*}
\caption{
Optical depths of the H$^{13}$CO$^+$ $J$=3$\rightarrow$2 lines toward L1448-mm (0,0) and (0,-10) 
derived from different isotopologue ratios for the ambient (amb.) and the 
precursor (prec.) components, and for different source morphologies.} 

\begin{center}
\begin{tabular}{ll|cc|cc}
\hline

\multicolumn{6}{c}{(0,0)} \\ \hline
 Ratio used&  Component  & $T_\text{B,rare}/T_\text{B,32}$& $\tau_{32}$(H$^{13}$CO$^+$)  & $T_\text{B,rare}/T_\text{B,32}$& $\tau_{32}$(H$^{13}$CO$^+$) \\ \hline
& &  \multicolumn{2}{|c|}{Source size $5''$}&\multicolumn{2}{|c}{Extended source}\\ \cline{3-6}
HC$^{18}$O$^+$/H$^{13}$CO$^+$&Amb.&$\leq 2.4$&...&$\leq0.9$&$\leq16$\\ 
&Prec.&$\leq0.25$&$\leq2.1$&$\leq0.09$&$\leq 0.59$$^{\mathrm{a}}$\\
H$^{13}$C$^{18}$O$^+$/H$^{13}$CO$^+$&Amb.&$\leq1.1$&...& $\leq 0.4$&$\leq290$\\
&Prec.&$\leq 0.11$&$\leq67$&$\leq0.04$&$\leq24$\\ \hline \hline

\multicolumn{6}{c}{(0,-10)} \\ \hline
 Ratio used&  Component & $T_\text{B,rare}/T_\text{B,32}$& $\tau_{32}$(H$^{13}$CO$^+$)  & $T_\text{B,rare}/T_\text{B,32}$& $\tau_{32}$(H$^{13}$CO$^+$) \\ \hline
& & \multicolumn{2}{|c|}{Source size $5''$}&\multicolumn{2}{|c}{Extended source}\\ \cline{3-6}
HC$^{18}$O$^+$/H$^{13}$CO$^+$&Amb.&$\leq3.1$&...& $\leq1.1$&...\\
&Prec.& $\leq0.85$&$\leq 16$ &$\leq0.3$&$\leq2.8$\\
H$^{13}$C$^{18}$O$^+$/H$^{13}$CO$^+$&Amb.&$\leq1.4$&...&$\leq0.5$&$\leq390$\\
&Prec. &$\leq0.38$&$\leq280$&$\leq0.1$&$\leq83$\\ \hline

\end{tabular}

\begin{list}{}{}
\item[$^{\mathrm{a}}$] This value of $\tau$(H$^{13}$CO$^+$) was estimated 
assuming $R$=$\frac{40}{577}$, rather than $R$=$\frac{69}{577}$ (see text for explanation).
\end{list}

\end{center}
\label{tab:opacities1}
\end{table*}%

%----------------------------------------------------------------------------------------

%TABLE 6 --------------------------------------------------------------------------------

\begin{table*}
\caption{
Optical depths of the HN$^{13}$C $J$=3$\rightarrow$2 line toward L1448-mm (0,0) and (0,-10) 
derived from different isotopologue ratios for the ambient (amb.) and the 
precursor (prec.) components, and for different source morphologies.}
\begin{center}
%\begin{tabular}{ll|cc|cc|cc}
\begin{tabular}{ll|cc|cc}
\hline 

\multicolumn{6}{c}{(0,0)} \\ \hline
 Ratio used& Component & $T_\text{B,rare}/T_\text{B,32}$& $\tau_{32}$(HN$^{13}$C)  & $T_\text{B,rare}/T_\text{B,32}$& $\tau_{32}$(HN$^{13}$C) \\ \hline
& &  \multicolumn{2}{|c|}{Source size $5''$}&\multicolumn{2}{|c}{Extended source}\\ \cline{3-6}
H$^{15}$NC/HN$^{13}$C&Amb.&$\leq0.74$&$\leq7.6$&$\leq0.3$&$\leq1.1$ \\
&Prec.&$\sim1.1$&...&$\sim0.4$&$\sim2.5$\\
H$^{15}$N$^{13}$C/HN$^{13}$C&Amb.&$\leq1.6$&...&$\leq 0.6$&$\leq320$\\
&Prec.&$\sim2.3$&...& $\sim0.8$&$\sim 700$\\\hline \hline

\multicolumn{6}{c}{(0,-10)} \\ \hline
Ratio used & Component & $T_\text{B,rare}/T_\text{B,32}$& $\tau_{32}$(HN$^{13}$C)  & $T_\text{B,rare}/T_\text{B,32}$& $\tau_{32}$(HN$^{13}$C) \\ \hline
& & \multicolumn{2}{|c|}{Source size $5''$}&\multicolumn{2}{|c}{Extended source}\\ \cline{3-6}
H$^{15}$N$^{13}$C/HN$^{13}$C&Amb.&$\sim2.2$&...&$\sim0.8$&$\sim610$\\
&Prec.& $\sim2.2$&...&$\sim0.8$&$\sim610$\\ \hline 

\end{tabular}

\end{center}
\label{tab:opacities2}
\end{table*}%
%-----------------------------------------------------------------------------------------

\section{Excitation differences between ion and neutral molecular fluids}
\label{section:excitation}

\subsection{Revisiting the ion and neutral molecular excitation analysis toward
L1448-mm: New collisional data and the effect of temperature and source size}
\label{coll_coeffs}

JS06 proposed that the over-excitation of the 
H$^{13}$CO$^{+}$ molecular ions observed in the precursor component 
toward L1448-mm (0,0) and  (0,-10) could be explained by an 
enhanced fractional abundance of electrons of $\sim 5 \times 10^{-5}$.
We need to check if this conclusion still holds when using the newly released collisional coefficients of SiO and
HNC with electrons  \citep{Varambhia09,Faure07}.

 In addition, JS06 assumed that the kinetic temperature of the gas was low [21~K, estimated from NH$_3$ observations \citep{Curiel99}], but in this section we also explore the possibility that the gas is  at higher temperatures, which is predicted by C-shock models.
Finally, we investigate the possibility that the source size is compact.

As the excitation model, we have used the LVG code of JS06, but using the  new collisional coefficients 
with electrons for SiO and HN$^{13}$C.
It should be noted that for any
isotopologues, de-excitation rates of the main species were employed, except
for H$^{13}$CO$^+$. For the latter, a proper calculation was performed and
minor differences (less than 5\%) were observed with respect to HCO$^+$ rate
coefficients. Excitation rates were derived from the detailed balance
principle. 

\subsubsection{Extended emission with low kinetic temperatures}
\label{ext}

The estimated H$_2$ densities and column densities required to produce the observed line intensities are shown in Table~\ref{cold}, assuming that excitation is due to collisions with H$_2$ only. 
 As in JS06, where possible we have used the $J$=4$\rightarrow$3 and $J$=3$\rightarrow$2 lines to derive the H$_2$ densities, as these transitions are more likely to trace the gas processed by the precursor compared to the lower excitation lines.  Where this information is not available, we have used the $J$=3$\rightarrow$2 and $J$=1$\rightarrow$0 lines to estimate the H$_2$ densities.  
The H$_2$ densities for all molecules are a factor of $\sim10$ times higher than those in Table~3 of JS06.  This difference mainly arises because JS06 used main beam temperatures to estimate the H$_2$ densities, which is equal to $T_\text{B}$ when the source size exactly fills the beam, whereas  in our case we have used $T_\text{A}^*$, appropriate for extended emission.

We find that an H$_2$ density of $\sim 10^6$~cm$^{-3}$ can explain the excitation of the neutral species, but the H$_2$ density required for H$^{13}$CO$^+$ is a factor of $\sim 5$ times larger toward (0,0), and $\sim 10$ times larger toward (0,-10),  as previously found by JS06.

 If we now assume an H$_2$ density of $1\times10^6$~cm$^{-3}$, fractional ionisations of 
$5\times10^{-4}$ and $7\times10^{-3}$ toward (0,0) and (0,-10), repectively,  are required to 
explain the line intensity ratios measured for H$^{13}$CO$^+$ in the precursor component.  
These enhanced electron abundances do not make a strong impact on the predicted intensities of SiO and HN$^{13}$C, since they differ by less than 30 per cent for SiO, and by a factor of 2 for HN$^{13}$C, with respect to the observed ones.  This shows that when using the new collisional coefficients of SiO and HN$^{13}$C in the calculations, the electron density enhancement in the precursor component is consistent with the measured intensities of both the ion and the neutral species.
The LVG model estimates the column densities and opacities of the H$^{13}$CO$^+$ $J$=3$\rightarrow$2 lines to be $N= 1\times10^{12}$~cm$^{-2}$ and $\tau_{32}=0.2$ toward (0,0) and $N= 3\times10^{11}$~cm$^{-3}$ and $\tau_{32}=0.05$ toward (0,-10).  These opacities are consistent  with the observed upper limits in Section~\ref{opacities_calc}.

\begin{table*}
\caption{H$_2$ densities and molecular column densities of 
H$^{13}$CO$^+$ and HN$^{13}$C derived assuming extended emission and a temperature of 21~K for all molecules, for the precursor component of L1448-mm. }
\begin{center}
\begin{tabular}{lcc}
\hline 
&  H$_2$ density (cm$^{-3}$)& Column density (cm$^{-2}$) \\ \cline{2-3} 
Molecule &  & \\
\hline
\multicolumn{3}{c}{(0,0)}\\ \hline
H$^{13}$CO$^+$   & $5.0\times10^6$ & $9.3\times10^{11}$ \\
SiO & $1.6\times10^6$ & $1.2\times10^{11}$ \\
HN$^{13}$C   & $\leq 1.3\times10^6$ & $3.7\times10^{11}$\\ 
H$^{13}$CN & $\leq4.8\times10^6$ & $3.3\times10^{11}$  \\ \hline
\multicolumn{3}{c}{(0,-10)}\\ \hline
H$^{13}$CO$^+$   & $3.7\times10^7$ & $2.6\times10^{11}$ \\
SiO & $3.5\times10^6$ & $1.4\times10^{11}$ \\
HN$^{13}$C   & $\leq 7.1\times10^5$ & $6.8\times10^{11}$\\ 
H$^{13}$CN & $\leq3.6\times10^6$ & $4.2\times10^{11}$  \\ \hline

\end{tabular}
\end{center}
\label{cold}
\end{table*}

\subsubsection{Extended emission with high kinetic temperatures}

Since the collisional coefficients for H$^{13}$CO$^+$ with electrons used so far in the LVG model are only available for transitions up to $J=5$, which lies at $\sim62.5$~K, for the LVG calculations at higher temperatures we use updated collisional coefficients for HCO$^+$ with electrons \citep{Faure01}, with data for transitions up to $J=20$, lying at $\sim900$~K.  To ensure that the LVG calculations are converged with respect to the 
rotational populations, we can only perform LVG calculations for temperatures $\lesssim 400$~K.

If the gas has a kinetic temperature of $100$~K, the estimated H$_2$ densities for the neutral species fall to $\sim 3\times 10^5$~cm$^{-3}$.  For H$^{13}$CO$^+$, the estimated H$_2$ densities are $\sim 4\times 10^5$~cm$^{-3}$ and $\sim 8\times 10^5$~cm$^{-3}$ toward (0,0) and (0,-10) respectively.  Assuming that the gas density is $3\times10^5$~cm$^{-3}$, the H$^{13}$CO$^+$ ions would  still require fractional ionisations of $3\times10^{-4}$ and $8\times10^{-4}$ toward (0,0) and (0,-10) respectively to fit the observations.

Performing a similar analysis at 200~K, 300~K and 400~K, we find that toward (0,0), the required H$_2$ density to fit the H$^{13}$CO$^+$  data is a factor of $1.5-2$ times larger than the H$_2$ density required for the neutral species, and toward (0,-10) this factor  is $2.5-4$.  The fractional ionisations required to fit the data assuming only one H$_2$ density, 
range from $3\times10^{-4}$ to  $2 \times10^{-3}$ toward (0,0), and from $8\times10^{-4}$ to  $4 \times10^{-3}$, toward (0,-10).
 Without the electron density enhancements, the H$^{13}$CO$^+$ $J$=3$\to$2 lines are underpredicted by $20-60$\%, and the H$^{13}$CO$^+$ $J$=4$\to$3 lines are underpredicted by $30-70$\%.
The estimated column densities and opacities of H$^{13}$CO$^+$ $J$=3$\rightarrow$2 are slightly lower than for the 21~K case above, with $N= 8-9\times10^{11}$~cm$^{-2}$ and $\tau_{32}=0.2$ toward (0,0) and $N= 2\times10^{11}$~cm$^{-2}$ and $\tau_{32}=0.03$ toward (0,-10).   Again these opacities are consistent with the upper limits derived from observations  (see Section$\,$4).

\begin{figure}
 \includegraphics[angle=0, width=0.5\textwidth]{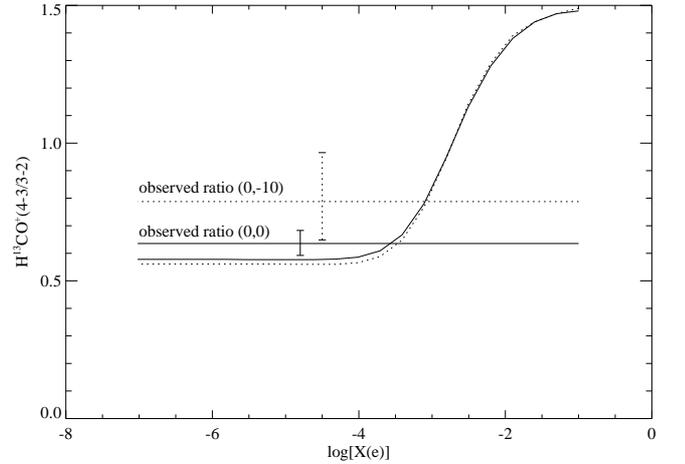}
 \caption{The H$^{13}$CO$^+$ $J$=4$\rightarrow$3$/J$=3$\rightarrow$2 ratio as predicted by the LVG model at 100~K toward L1448-mm (0,0) (solid curve) and L1448-mm (0,-10) (dotted curve), against the electron fractional abundance.  
The observed $J$=4$\rightarrow$3$/J$=3$\rightarrow$2 ratios of the pre-cursor components towards both positions are plotted, and the error bars correspond to the $1\sigma$ errors in the observations.
}
 \label{e-effect}
 \end{figure}

Note that toward (0,0), even though the neutral species require an H$_2$ density of {\it only} $1.5-2$ times less than  that for H$^{13}$CO$^+$, the estimated fractional ionisation is very high, up to four orders of magnitude higher than the typical fractional ionisation in dark clouds ($\sim10^{-7}$).  This is because the electrons do not significantly affect the excitation of H$^{13}$CO$^+$ until they reach a fractional abundance of  a few $\times10^{-5}$.  
 This is demonstrated by Figure~\ref{e-effect}, where we show the H$^{13}$CO$^+$ $J$=4$\rightarrow$3$/J$=3$\rightarrow$2 ratio as a function of the electron fractional abundance
predicted by the LVG model, for an H$_2$ density of $3\times10^{5}$~cm$^{-3}$, and a temperature of 100~K.  
For each position we have used the column densities estimated above, although Figure~\ref{e-effect} shows that variations in the column density have a minimal
 impact on the predicted $J$=4$\rightarrow$3$/J$=3$\rightarrow$2 ratios.

 We note that considering the $1\sigma$ errors in the observations, toward (0,0) the lower limit of  the observed H$^{13}$CO$^+$ $J$=4$\rightarrow$3$/J$=3$\rightarrow$2 ratio is approximately equal to that predicted by the LVG model for low electron densities ($X$(e)$\leq10^{-5}$) for temperatures of 100--400~K.  It is therefore possible that there is no need for an electron density enhancement toward (0,0) for these temperatures.  However, toward (0,-10), as demonstrated by the error bars plotted in Figure~\ref{e-effect}, the need for an electron density enhancement to explain the observations is more compelling; the lower limit of the observed $J$=4$\rightarrow$3$/J$=3$\rightarrow$2 ratio is 13-24 per cent larger than that predicted by the LVG model for low electron densities for these high temperatures [compared to within 0.2--12 per cent toward (0,0)].

 %  effect of electrons figure-------------------------------------------------------------
 %----------------------------------------------------------------------------------------------

\subsubsection{Compact emission}
\label{compact}
Assuming a source size of $5''$, converting the $T_\text{A}^*$s to brightness temperatures, as described in Section~\ref{opacities_calc}, gives brightness temperatures of $40-50$~K for the H$^{13}$CO$^+$ $J$=1$\rightarrow$0 lines, and $15-30$~K for the HN$^{13}$C $J$=1$\rightarrow$0 lines.  Therefore, if the emission is compact, the kinetic  temperatures of the gas must be $\geq50$~K.  
 The brightness temperatures for the H$^{13}$CO$^+$ $J$=3$\rightarrow$2 and $J$=4$\rightarrow$3 lines are only $20-25$~K toward (0,0) and $\sim7$~K toward (0,-10). For the HN$^{13}$C $J$=3$\rightarrow$2 lines, their brightness temperatures are $\lesssim3$~K toward both positions.
For gas temperatures $>50$~K, our LVG model cannot fit the HN$^{13}$C intensities toward (0,0) and (0,-10), assuming that  the molecular excitation is  produced by collisions with  only H$_2$.  This is because we cannot fit the extremely low values of the $J$=3$\rightarrow$2/$J$=1$\rightarrow$0 brightness temperature ratios  for HN$^{13}$C. 
In fact, the LVG model cannot  reproduce the HN$^{13}$C intensities even for kinetic temperatures of $15-30$~K.  
 There is a similar problem for H$^{13}$CO$^+$; whilst it is possible to find an H$_2$ density and H$^{13}$CO$^+$ column density to fit the $J$=4$\rightarrow$3 and $J$=3$\rightarrow$2 observations for kinetic temperatures $\geq50$~K, these fits cannot match the extremely high intensities of the $J$=1$\rightarrow$0 lines, and it is impossible to fit the extremely low $J$=3$\rightarrow$2$/J$=1$\rightarrow$0 line intensity ratio.  Including an electron density enhancement increases the $J$=3$\rightarrow$2/$J$=1$\rightarrow$0, making the difference between the model and observed values even larger.
Furthermore, observations of NH$_3$ toward L1448-mm indicate that the neutral gas has a temperature of $\sim20$~K \citep{Curiel99}, lower than the minimum temperature of $\sim50$~K required for compact emission.  
The observational evidence and our modelling results are therefore not consistent with compact emission.

\subsection{Analysis of new positions around the L1448-mm source}
\label{new-pos-obs}

In this section, we present the analysis of the molecular data  
toward the newly observed positions L1448-mm (0,20), (20,0) and (-20,0). We
use again the LVG approximation as the molecular excitation model. The ion 
and neutral low- and high-$J$ lines of H$^{13}$CO$^+$ and HN$^{13}$C 
measured toward these offsets, are needed to constrain not only the spatial 
extent of the emission arising from the magnetic precursor, but also the
extent of the electron density enhancement associated with this early  shock stage. 
As suggested by JS06 for the L1448-mm (0,0) and (0,-10) positions, 
if an electron density enhancement occurs in the precursor, 
the high-$J$ rotational lines of H$^{13}$CO$^+$ 
are expected to be significantly brighter than those of HN$^{13}$C. 

If the emission is optically thin (see Section~\ref{opacities_calc}), the line 
intensity ratio between different transitions is directly related to the excitation 
temperature of the molecular species. In Table~\ref{ratios}, we show the 
$J$=3$\rightarrow$2/1$\rightarrow$0 integrated intensity ratios for H$^{13}$CO$^+$ and 
HN$^{13}$C toward all positions measured in L1448-mm.

For H$^{13}$CO$^{+}$, we find that the $J$=3$\rightarrow$2/1$\rightarrow$0 ratio is 
significantly larger in the precursor component than in the ambient gas 
(by more than a factor of 3), {\it only} toward L1448-mm (0,0) and
(0,-10). For HN$^{13}$C, this ratio is similar (i.e. differs by less than 
a factor of 3) in both velocity components toward all positions, suggesting
that the over-excitation of H$^{13}$CO$^{+}$ does not show a large-scale morphology
around the L1448-mm core. This clearly contrasts with the idea that the anomalous
excitation of H$^{13}$CO$^+$ is produced by a diffusion/scattering effect 
(Frerking et al. 1979; Cernicharo et al. 1984).

This can also be qualitatively shown by deriving the H$_2$  densities 
required to explain the line intensity ratios shown in Table~\ref{ratios}, 
as in Section~\ref{ext}. 
 We assume that the emission is spatially extended and consider a kinetic temperature of 21~K.
The estimated H$_2$
densities and molecular column densities are shown in Table~\ref{LVG}.

We see that the intensities of the low- and high-$J$ 
H$^{13}$CO$^+$ lines arising from the precursor component toward 
L1448-mm (0,20), (20,0) and (-20,0),  
can be explained by an H$_2$ density that differs only
by a factor of 2 with respect to that of the ambient gas, 
 and that is less than the upper limits of the H$_2$ denstities derived from
HN$^{13}$C within the precursor component.

These results suggest that the ion H$^{13}$CO$^{+}$ fluid does not show any evidence
of over-excitation with respect to the neutral fluid toward the newly observed 
offsets around L1448-mm. This implies that molecular excitation 
by collisions {\it only} with H$_2$, can explain the intensities of the low- and 
high-$J$ lines of H$^{13}$CO$^{+}$ and HN$^{13}$C measured toward these
new positions. The detected over-excitation of the ions 
toward L1448-mm (0,0) and (0,-10) seems to be restricted to a particular 
region located in the southern red-shifted lobe 
of this outflow. This is the expected behavior since very young shocks have been 
reported by \cite{Girart01} toward these positions.

%TABLE 10-----------------------------------------------------------------------------------

\begin{table*}
\caption{$J$=3$\rightarrow$2/1$\rightarrow$0 integrated intensity ratios. Data were taken 
from JS06 and this paper.}
\begin{center}
\begin{tabular}{lcccccc}
\hline 
&  \multicolumn{2}{c} {(0,20)}&   \multicolumn{2}{c} {(20,0)}& \multicolumn{2}{c} {(-20,0)} \\ \cline{2-3} \cline{4-5} \cline{6-7}
  Molecule   &  Ambient   &   Precursor &   Ambient  &    Precursor  &    Ambient     &  Precursor\\       
\hline
  H$^{13}$CO$^+$ &      0.2  & 0.5 & $\leq 0.2$ & 0.4  &  $\leq 0.2$ &   0.09 \\
HN$^{13}$C &    $\leq 0.1$ &  $\leq 0.3$ &$\leq 0.1$& $\leq 0.1$&   $\leq 0.1$ & $\leq 0.3$\\ \hline

\hline \hline
&  \multicolumn{2}{c} {(0,0)}&   \multicolumn{2}{c} {(0,-10)}& \multicolumn{2}{c} {(0,-20)} \\ \cline{2-3} \cline{4-5} \cline{6-7}
 Molecule   &  Ambient   &   Precursor &   Ambient  &    Precursor  &    Ambient     &  Precursor\\       
\hline
  H$^{13}$CO$^+$ &  0.5 & 1.6 & $\leq0.1$ & 0.6  & $\leq0.5$  & 0.2   \\
HN$^{13}$C & 0.4  & $\leq0.2$  & $\leq0.2$ &$\leq0.1$  & $\leq0.2$ &$\leq0.3$ \\ \hline

\end{tabular}
\end{center}
\label{ratios}
\end{table*}
%------------------------------------------------------------------------------------------

%TABLE 11----------------------------------------------------------------------------------

\begin{table*}
\caption{H$_2$ volume densities and molecular column densities of 
H$^{13}$CO$^+$ and HN$^{13}$C derived assuming extended emission for both 
the ambient and the precursor components of L1448-mm.
}
\begin{center}
\begin{tabular}{lcccc}
\hline 
& \multicolumn{2}{c} { H$_2$ density (cm$^{-3}$)}&\multicolumn{2}{c} { Column density (cm$^{-2}$)} \\ \cline{2-3} \cline{4-5}
 Molecule & Ambient & Precursor & Ambient & Precursor\\
\hline
\multicolumn{5}{c}{(0,20)}\\ \hline
H$^{13}$CO$^+$   &   $1.4\times10^5$   & $3.5\times10^5$   &$4.8\times10^{11}$  & $4.0\times10^{11}$  \\
HN$^{13}$C   & $\leq 1.9\times10^5$    &  $\leq 9.2\times10^5$  &  $\leq 6.4\times10^{11}$  & $\leq 1.5\times10^{11}$ \\ \hline
\multicolumn{5}{c}{(20,0)}\\ \hline
H$^{13}$CO+   & $\leq 8.8\times10^4$ &  $1.8\times10^5$    & $\leq 5.1\times10^{11}$    &  $4.3\times10^{11}$  \\
HN$^{13}$C   & $\leq 1.8\times 10^5$ & $\leq 3.3\times10^5$  &  $\leq 7.2\times10^{11}$   &  $\leq 4.1\times10^{11}$ \\ \hline
\multicolumn{5}{c}{(-20,0)}\\ \hline
H$^{13}$CO$^+$    &  $\leq 8.7\times10^4$     & $7.1\times 10^4$ & $\leq 3.5\times10^{11}$   & $6.4\times10^{11}$\\
HN$^{13}$C   &   $1.2\times10^5$  & $\leq 6.1\times10^5$    & $1.4\times10^{12}$   & $\leq 2.0\times10^{11}$ \\ \hline

\end{tabular}
\end{center}
\label{LVG}
\end{table*}
%------------------------------------------------------------------------------------------------------------------------------------------

\section{Discussion and conclusions}
\label{obs_conclusions}

Anomalies in the intensities of the hyperfine components of the 
strongly polar HCN molecule  were reported on large-scales toward the 
TMC-1 molecular dark cloud by \cite{Cernicharo84}. These authors showed that these anomalies 
are  due to large opacity effects and are produced by the re-emission 
of radiation coming from the densest part of the core, by a more extended and less
dense envelope. If this applies to the L1448-mm case, then we should expect to detect 
such differences in the excitation of other strongly polar molecular species such as
HCO$^{+}$ and HNC, on large-scales around the L1448-mm core. 
 In Section~\ref{opacities_calc}, from isotopologue observations we have shown that 
the precursor component of H$^{13}$CO$^+$ is optically thin toward (0,0), and if the emission 
is extended, this component is also optically thin toward (0,-10).
This implies that the over-excitation of H$^{13}$CO$^+$ in the precursor toward these 
positions cannot be due to a large-scale optical depth/scattering effect. 

In view of the recent release of new collisional coefficients of SiO and HNC with
electrons \citep{Faure07,Varambhia09}, we have re-analysed 
the multi molecular line observations toward L1448-mm (0,0) and (0,-10) of JS06, but taking into account different 
possibilities for the source 
size extent  and temperature. 
 Assuming extended emission and kinetic temperatures of up to 400~K, and considering that the molecular excitation is due to collisions with only H$_2$, a higher H$_2$ density is required to match the observations of H$^{13}$CO$^+$ than that needed for the neutral species. 
For temperatures of $100-400$~K, toward L1448-mm (0,0) it is possible that, considering the $1\sigma$ errors in the observations, the observations can be explained by a single H$_2$ density for the ions and the neutrals, but toward (0,-10) these errors are not enough to explain the difference in the required H$_2$ densities.
This leads to the need for an extra excitation mechanism for H$^{13}$CO$^+$ in the precursor component of L1448-mm.  This extra excitation can be provided by an electron density enhancement of up to a factor of $\sim10^4$.  
 As in JS06, the derived electron density enhancement exceeds those predicted by MHD shock modelling by up to two orders of magnitude \citep{Flower96,Flower03}.  However, simply considering the $1\sigma$ errors in the observations  can reduce the electron density estimates by an order of magnitude.  Due to these very large uncertainties, we can conclude that an electron density enhancement can explain the observations, but we cannot quantify it with an accuracy better than one order of magnitude.

If the emission is compact, the gas kinetic temparture must be $\gtrsim50$~K.  However, we find that the HN$^{13}$C and H$^{13}$CO$^+$ lines cannot be fitted in this case, so we conclude that the emitting region associated with the precursor must be $>5''$.

The observations toward the L1448-mm (0,20), (20,0) and (-20,0) positions, 
do not show evidence for an over-excitation of H$^{13}$CO$^{+}$ with respect to
the neutral HN$^{13}$C molecules in the precursor component of L1448-mm. 
This suggests that the extra-mechanism 
responsible for the over-excitation of the ions has its origin in local
phenomena likely linked to the recent interaction of very young C-shocks.
\cite{Girart01} have indeed reported very young shocks 
(with a dynamical age of $\leq 90$~yr) toward the southern position L1448-mm 
(0,-10), making the probability to detect the magnetic precursor of C-shocks 
not negligible. Since our results show that the over-excitation of ions is 
confined to this region, there seems to be strong evidence to suggest that this 
over-excitation is produced by the electron density enhancement at 
the magnetic precursor stage of young shocks. Furthermore, the confinement of the 
over-excitation of the ions also implies that this effect is unlikely to be 
caused by external UV illumination, as it should be observed tracing the inner
regions of the outflow cavity, and on larger spatial scales. 

In order to fully understand the extent of the magnetic precursor of L1448-mm, 
high angular resolution observations of the high-$J$ transitions of H$^{13}$CO$^+$, 
HN$^{13}$C and SiO carried out with interferometers such as the Submillimeter 
Array (SMA), are therefore strongly required to measure the source sizes of 
this molecular emitting regions and to confirm whether or not this emission 
is optically thin. 
 Observations of very high-$J$ ($J>10$) transitions are also required to verify whether or not the gas has reached the high temperatures predicted by C-shock models, as these lines should be detected for such temperatures. 
It is also necessary to search for signs of the electron 
density enhancement toward other young outflows where narrow SiO has been detected, 
such as L1448-IRS3, NGC1333-IRS4 and NGC1333-IRS2, in order to investigate whether 
the electron density enhancement is a common phenomenon in this type of objects, 
and if it is correlated with the detection of very narrow SiO emission centred 
at ambient cloud velocities. 

\section{Acknowledgements}
This work has been partially supported by the Spanish Ministerio de Ciencia
e Innovaci\'on under project ESP2007-65812-C02-01, and by the Comunidad de
Madrid Government under PRICIT project S-0505/ESP-0237 (ASTROCAM).

{}
%\listofobjects

\end{document}